\documentclass[aps,twocolumn,prl,superscriptaddress,amsmath,amssymb,groupedaddress]{revtex4-1}

\usepackage{dcolumn}
\usepackage{bm}

\ifx\pdfoutput\undefined
\usepackage[dvipdfmx]{graphicx}
\usepackage[dvipdfmx]{hyperref}
\else
\usepackage{graphicx}
\usepackage{hyperref}
\fi

\usepackage{color}
\usepackage{ulem}

\begin{document}

\title{
Majorana-Magnon Crossover by a Magnetic Field in the Kitaev Model:\\
Continuous-Time Quantum Monte Carlo Study
}

\author{Junki Yoshitake$^1$, Joji Nasu$^2$, Yasuyuki Kato$^1$, and Yukitoshi Motome$^1$}
\affiliation{$^1$Department of Applied Physics, University of Tokyo, Hongo 7-3-1, Bunkyo, Tokyo 113-8656, Japan\\
$^2$Department of Physics, Yokohama National University, Hodogaya, Yokohama 240-8501, Japan
}

\date{\today}

\begin{abstract}
Kitaev quantum spin liquids host Majorana fermions via the fractionalization of spins. 
In a magnetic field, the Majorana fermions were predicted to comprise a topological state, which has attracted great attention by the discovery of the half-quantized thermal Hall conductivity. 
Nevertheless, a reliable theory remains elusive for the field effect, especially at finite temperature. 
Here we present unbiased large-scale numerical results for the Kitaev model in a wide range of magnetic field and temperature. 
We find that the unconventional paramagnetic region showing fractional spin dynamics extends at finite temperature, far beyond the field range where the topological state is expected at zero temperature. 
Our results show the confinement-deconfinement behavior between the fractional Majorana excitations and the conventional magnons. 
\end{abstract}

\maketitle


Conventional magnets exhibit magnetic long-range orders at low temperature ($T$). 
The low-lying excitations are described by bosonic quasiparticles called magnons. 
Such a magnon picture, however, breaks down when the magnetic order is suppressed down to the lowest $T$. 
Of particular interest is a quantum disordered state called the quantum spin liquid (QSL)~\cite{Anderson1973,Balents2010,Zhou2017,Savary2017}. 
In the QSL, the excitations are described by quasiparticles emergent from the fractionalization of spins~\cite{Read1989,Read1991,Wen1991,Senthil2000,Oshikawa2006}. 
The transition from magnons to fractional quasiparticles is one of the confinement-deconfinement problems, which have long been studied in not only condensed matter physics but also high-energy physics. 

Understanding of the deconfined fractional quasiparticles has been significantly gained by the intense research on the Kitaev QSL~\cite{Kitaev2006}.
The Kitaev QSL is realized in the ground state exactly obtained for the Kitaev model by representing the spins by Majorana fermions, which offers an explicit form of the fractionalization~\cite{Kitaev2006}. 
Since a class of Mott insulators with strong spin-orbit coupling can be described by the Kitaev model~\cite{Jackeli2009}, extensive theoretical and experimental studies have been performed for exploring the Kitaev QSL and capturing the signatures of the fractional excitations~\cite{Trebst2017,Winter2017a,Takagi2019}. 

Amongst others, the effect of a magnetic field has recently attracted great attention. 
The perturbation theory with respect to the field strength predicts a gapped topological ground state where the Majorana fermions turn into non-Abelian anyons~\cite{Kitaev2006}. 
Meanwhile, for a candidate material $\alpha$-RuCl$_3$, unconventional behaviors were reported as the signature of the fractional excitations when the parasitic magnetic order is destabilized by the magnetic field~\cite{Baek2017,Zheng2017,Wang2017,Ponomaryov2017,Banerjee2018,Jansa2018,Nagai2018,Wellm2018}. 
While it is still in debate whether this field-induced disordered state is the Kitaev QSL, the recent discovery of the half-quantized thermal Hall conductivity~\cite{Kasahara2018} is driving excitement in the confinement-deconfinement phenomenon in this system.

In the study of the field effect, a major obstacle is that there is less reliable theory for the Kitaev model in the magnetic field, especially at finite $T$. 
At zero field, although the exact solution is limited to the ground state~\cite{Kitaev2006}, the finite-$T$ properties have been studied by numerical techniques based on the Majorana representation~\cite{Nasu2014,Nasu2015,Yoshitake2016,Yoshitake2017a,Yoshitake2017b,Mishchenko2017,Dwivedi2018,Self2019}. 
The magnetic field also breaks the exact solvability, and furthermore, it spoils the applicability of the finite-$T$ numerical methods. 
Although the exact diagonalization was performed for small clusters~\cite{Yadav2016,Winter2017b,Winter2018,Hickey2019}, higher resolution in both energy and momenta is desired for clarifying the confinement-deconfinement behavior in the magnetic field. 

In this Letter, we present our numerical results on the Kitaev model in the magnetic field ($h$) at finite $T$, obtained by a continuous-time quantum Monte Carlo (CTQMC) method. 
By adopting a cluster-type expansion in the spin representation, we successfully reach up to $\sim 100$ spins down to $T \simeq 0.04 J$ in a wide range of $h$ ($J$ is the Kitaev exchange coupling). 
Calculating the thermodynamics and spin dynamics by using this technique, we unveil that the unconventional paramagnetic (PM) region hosting fractional excitations is extended in a wide range of $h$ and $T$. 
We also clarify how the spin dynamics changes from the fractional region to the forced ferromagnetic (FM) region while increasing $h$. 
Our unbiased numerical results with high resolution provide a firm ground for understanding of the field effects on the Kitaev magnets.


\begin{figure}[t]
    \includegraphics[width=\columnwidth,clip]{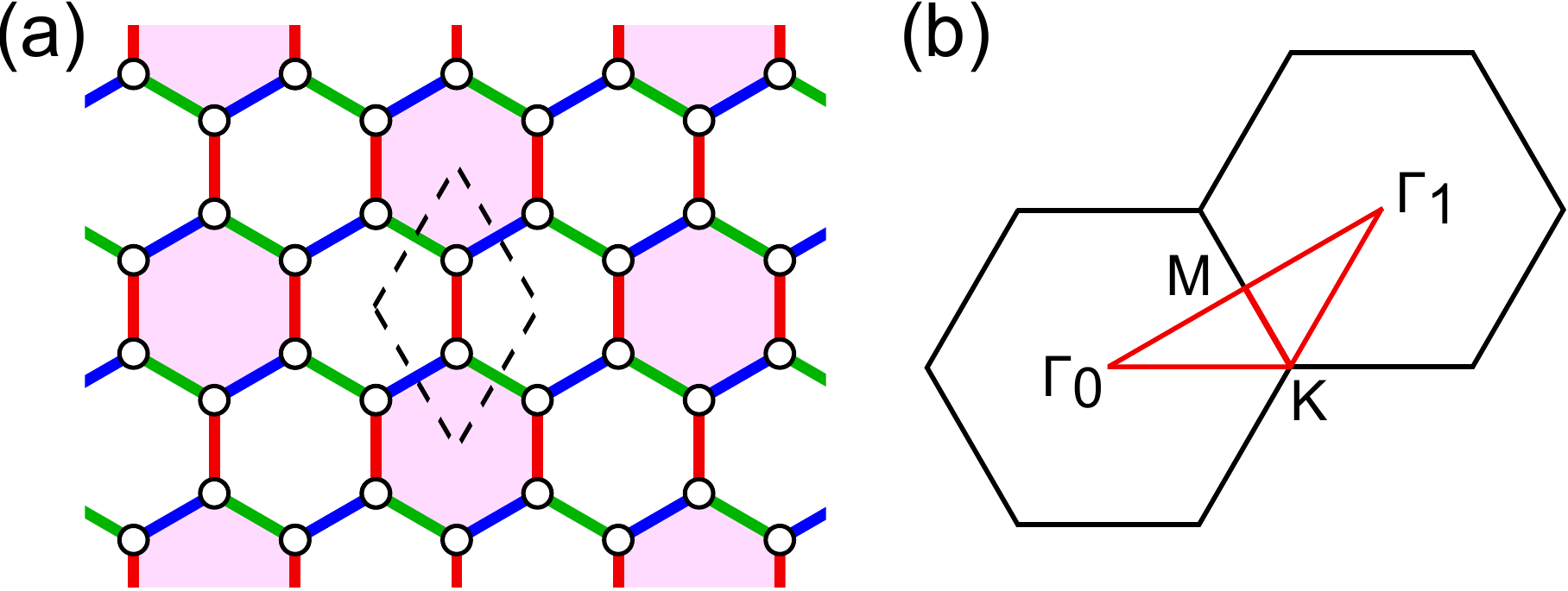}
    \caption{ \label{fig:fig1}
    (a) Schematic picture of the Kitaev model on the honeycomb structure. 
	The blue, green, and red bonds represent the $x$, $y$, and $z$ bonds in Eq.~(\ref{eq:H}). 
    The colored hexagons denote the unperturbed parts in the cluster-based CTQMC method. 
    The dashed diamond represents the unit cell. 
	(b) The first and second Brillouin zones centered at the $\Gamma_0$ and $\Gamma_1$ points, respectively. 
    The red lines represent the symmetric lines used in Fig.~\ref{fig:fig3}. 
    }
\end{figure}

The Kitaev model in the magnetic field is defined by the Hamiltonian~\cite{Kitaev2006} 
\begin{equation}
{\mathcal H} = - \sum_{\mu=x,y,z} J_\mu \sum_{\langle j,j' \rangle_\mu} S_j^\mu S_{j'}^\mu 
- \sum_j \mathbf{h} \cdot \mathbf{S}_j,
\label{eq:H} 
\end{equation}
where $\mathbf{S}_j = (S_j^x,S_j^y,S_j^z)$ represents the spin-1/2 operator at site $j$, $J_\mu$ denotes the exchange coupling for the nearest-neighbor (NN) spin pairs on the $\mu$ bonds, $\langle j,j' \rangle_\mu$ [see Fig.~\ref{fig:fig1}(a)], and $\mathbf{h} = (h^x,h^y,h^z)$ represents the magnetic field. 

Kitaev introduced a Majorana fermion representation of the model in Eq.~(\ref{eq:H}), where the spin-1/2 operator $\mathbf{S}_j$ is expressed by four Majorana fermion operators $c_j$ and $b_j^\mu$~\cite{Kitaev2006}. 
The Hamiltonian is rewritten into 
\begin{equation}
\tilde{\mathcal H} = i \sum_\mu \frac{J_\mu}{4} \sum_{\langle j,j' \rangle_\mu} u_{jj'}^\mu c_j c_{j'} 
- i \sum_j \sum_\mu \frac{h^\mu}{2} b_j^\mu c_j, 
\label{eq:H_Maj}
\end{equation}
where $u_{jj'}^\mu = i b_j^\mu b_{j'}^\mu$. 
At zero field, the bond variables $u_{jj'}^\mu$ are the $Z_2$ conserved quantities taking $\pm 1$. 
They are related with the $Z_2$ fluxes defined on each hexagonal plaquette $p$ as 
$W_p = \prod_{\langle j,j' \rangle_\mu \in p} u_{jj'}^\mu$, 
where the product is taken for the six sides of the hexagon $p$~\cite{Kitaev2006}. 
In this zero-field case, the ground state is exactly obtained as the flux-free state where all $W_p=1$, and the excitations from the ground state are described by the itinerant Majorana fermions $\{ c_j \}$ and the localized $Z_2$ fluxes $\{ W_p \}$. 
Thus, the excitations of the system are fractionalized into two types of quasiparticles. 
When switching on the magnetic field, however, the two quasiparticles are hybridized through the last term in Eq.~(\ref{eq:H_Maj}), and the exact solution has not been available thus far. 

We study the finite-$T$ properties of the Kitaev model under the magnetic field by using a CTQMC method. 
In the previous studies, the authors developed the QMC, cluster dynamical mean-field theory, and CTQMC methods based on a different type of the Majorana representation from Eq.~(\ref{eq:H_Maj})~\cite{Nasu2014,Nasu2015,Yoshitake2016,Yoshitake2017a,Yoshitake2017b}. 
The methods are, however, limited to zero magnetic field, as the Zeeman coupling term breaks the conservation of $W_p$ on which the methods rely. 
In the present study, we therefore develop another CTQMC method in the original spin representation in Eq.~(\ref{eq:H}). 
We perform a perturbation expansion~\cite{Rubtsov2005,Werner2006} by dividing the Hamiltonian in Eq.~(\ref{eq:H}) into two parts: the unperturbed one defined on the colored hexagons in Fig.~\ref{fig:fig1}(a) and the perturbed one for the bonds connecting the hexagons. 
The Zeeman term is included in the unperturbed part. 
This hexagonal clustering enables us to compute any spin correlations within the unperturbed Hamiltonian, which is required in the expansion type CTQMC simulation~\cite{note1}. 

While our method is applicable to any $J_\mu$ and any direction of $\mathbf{h}$, we here focus on the case with the isotropic FM coupling $J_x=J_y=J_z=J>0$ (we set $J=1$) and the magnetic field along the [111] direction $\mathbf{h} = \frac{1}{\sqrt{3}}(h,h,h)$~\cite{note4}. 
We show the benchmark for a small system size in the Supplemental Material~\cite{supp}. 
In the CTQMC calculations, the negative sign problem becomes severe at low $T$, but we can keep the precision of the simulations down to $T \simeq 0.04$ and up to the system with $N=96$ spins ($4 \times 4$ hexagons under periodic boundary conditions); we confirm that the finite-size effect is negligibly small~\cite{supp}. 
For the computational details of the CTQMC calculations, see the Supplemental Material~\cite{supp}.


\begin{figure}[t]
    \begin{center}
    \includegraphics[width=0.85\columnwidth,clip]{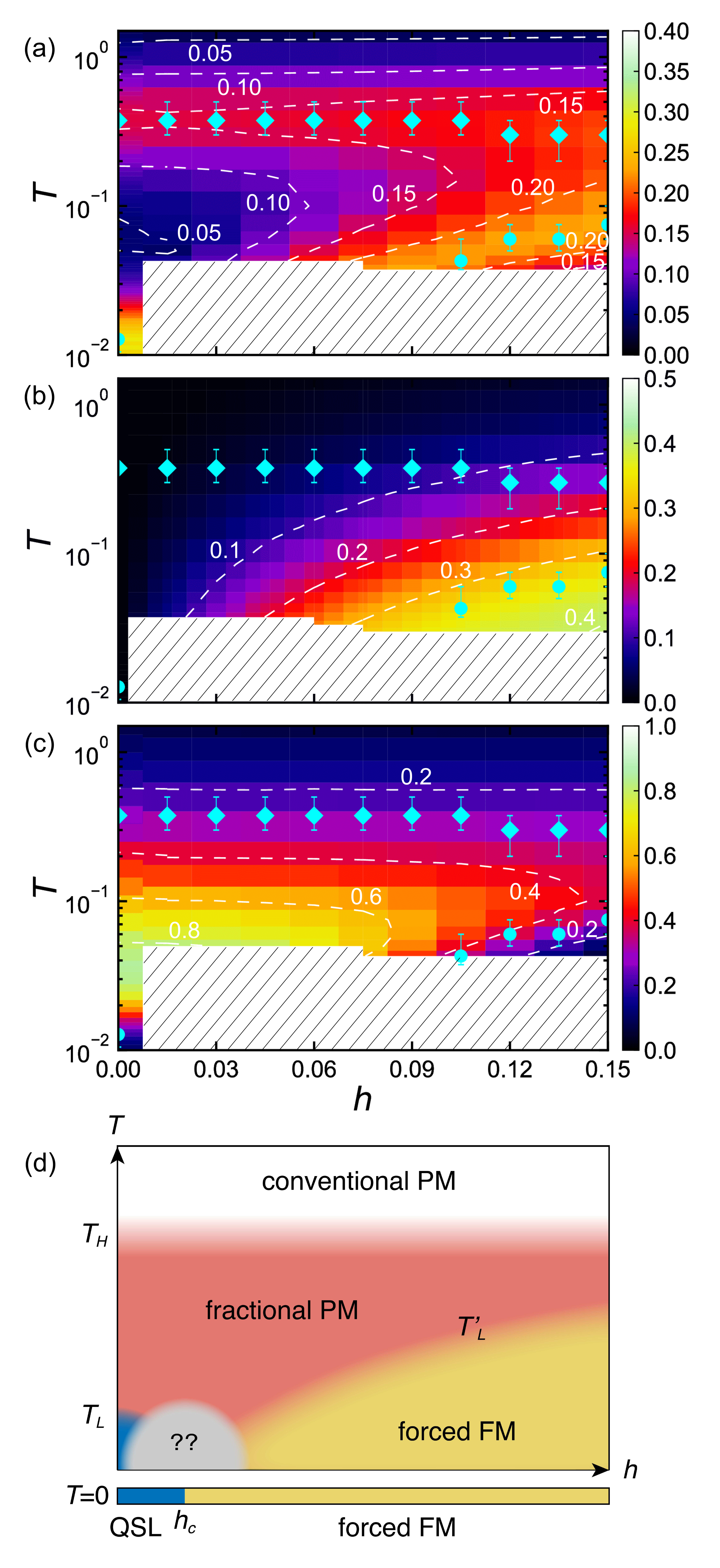}
    \caption{ \label{fig:fig2}
    Contour plots of (a) the specific heat per site, $C_v$, and (b) the magnetization per site, $M$, (c) the NN-site component of the NMR relaxation rate $1/T_1$ in the plane of the [111] field $h$ and temperature $T$~\cite{supp}.
	The crossover temperatures $T_H$, $T_L$, and $T'_L$ are plotted, which are estimated from the peaks in $C_v$~\cite{note3}. 
    The data at $h=0$ are taken from Ref.~\cite{Nasu2017} for (a) and Ref.~\cite{Yoshitake2017b} for (c).
    The hatched areas are the regions with severe negative sign. 
    (d) Schematic phase diagram. 
    The nature of the gray region is unexplored in the present simulation. 
    }
    \end{center}
\end{figure}

Figure~\ref{fig:fig2}(a) shows the $T$ and $h$ dependence of the specific heat per site, $C_v$, obtained by the cluster-based CTQMC simulation. 
The data at $h=0$ are taken from the previous QMC results~\cite{Nasu2017}, which could access down to below $T =0.01$ because of the absence of the negative sign problem~\cite{Nasu2015}. 
At $h=0$, $C_v$ shows two peaks at $T_H \simeq 0.375$ and $T_L \simeq 0.012$, as plotted in Fig.~\ref{fig:fig2}(a) by the diamond and circle, respectively. 
These are both crossovers; $T_H$ and $T_L$ roughly correspond to the energy scales for the center of mass of the itinerant Majorana bands and the $Z_2$ flux gap, respectively~\cite{supp}. 
Thus, the crossovers define the thermal fractionalization of spins~\cite{Nasu2015}; the PM state for $T_L \lesssim T \lesssim T_H$, dubbed the fractional PM state, is distinguished from the conventional PM state for $T \gtrsim T_H$ and the asymptotic QSL state for $T \lesssim T_{\rm L}$~\cite{Nasu2015}. 
For nonzero $h$, we find that $T_H$ remains at almost the same $T$. 
This suggests that the center of mass of the Majorana band is not changed significantly, even beyond the perturbation regime $h\ll 1$~\cite{supp}. 
Meanwhile, although the low-$T$ data for $h>0$ are not available due to the negative sign problem, the low-$T$ peak of $C_v$ reappears and shifts to higher $T$ in the high-$h$ region. 
We denote this temperature $T'_L$. 
Below this $T'_L$, as shown in Fig.~\ref{fig:fig2}(b), the magnetization per site, $M$, rapidly increases and becomes larger than $\sim 60$~\% of the full saturation. 
This indicates that $T'_L$ defines a crossover to the forced FM state, while $T_L$ is to the asymptotic QSL. 
Indeed, $T'_L$ scales well with the Zeeman splitting energy in the high-$h$ region~\cite{supp}. 
Thus, the results suggest that the fractional PM region extends to a wide region of $h$ for $T'_L \lesssim T \lesssim T_H$. 

The extended fractional PM region is also confirmed by the NMR relaxation rate $1/T_1$, which signals the dynamical spin fluctuations, as plotted in Fig.~\ref{fig:fig2}(c). 
At $h=0$, we plot the data obtained in the previous QMC+CTQMC study~\cite{Yoshitake2017b}: $1/T_1$ is enhanced below $T_H$ but suppressed below $T_L$ after showing a peak slightly above $T_L$. The enhanced $1/T_1$ is a good measure of the fractional spin dynamics in the fractional PM state~\cite{Yoshitake2016,Yoshitake2017a,Yoshitake2017b}. 
The result for $h>0$ in Fig.~\ref{fig:fig2}(c) indicates that the enhancement of $1/T_1$ survives in the region of $T'_L \lesssim T \lesssim T_H$, which supports the fractional PM state in the wide $h$-$T$ region. 
We note that the $h$-$T$ dependences of $1/T_1$ are qualitatively consistent with recent experiments~\cite{Baek2017,Zheng2017,Jansa2018,Nagai2018,supp}.

These results are summarized in the schematic phase diagram shown in Fig.~\ref{fig:fig2}(d). 
According to the previous study at $T=0$, the transition from the topological QSL state to the forced FM state occurs at the critical field $h_c \simeq 0.018$~\cite{Jiang2011}. 
Thus, although the data are lacked in low-$T$ and low-$h$ regions, our results indicate that the fractional PM state extends far beyond $h_c$ at finite $T$; the signatures of the fractional excitations can be captured not only in the low-$h$ region for $0 \leq h \lesssim h_c$ but also in the wide region of the fractional PM state extending up to $\sim 10h_c$. 

\begin{figure*}[t]
    \includegraphics[width=2\columnwidth,clip]{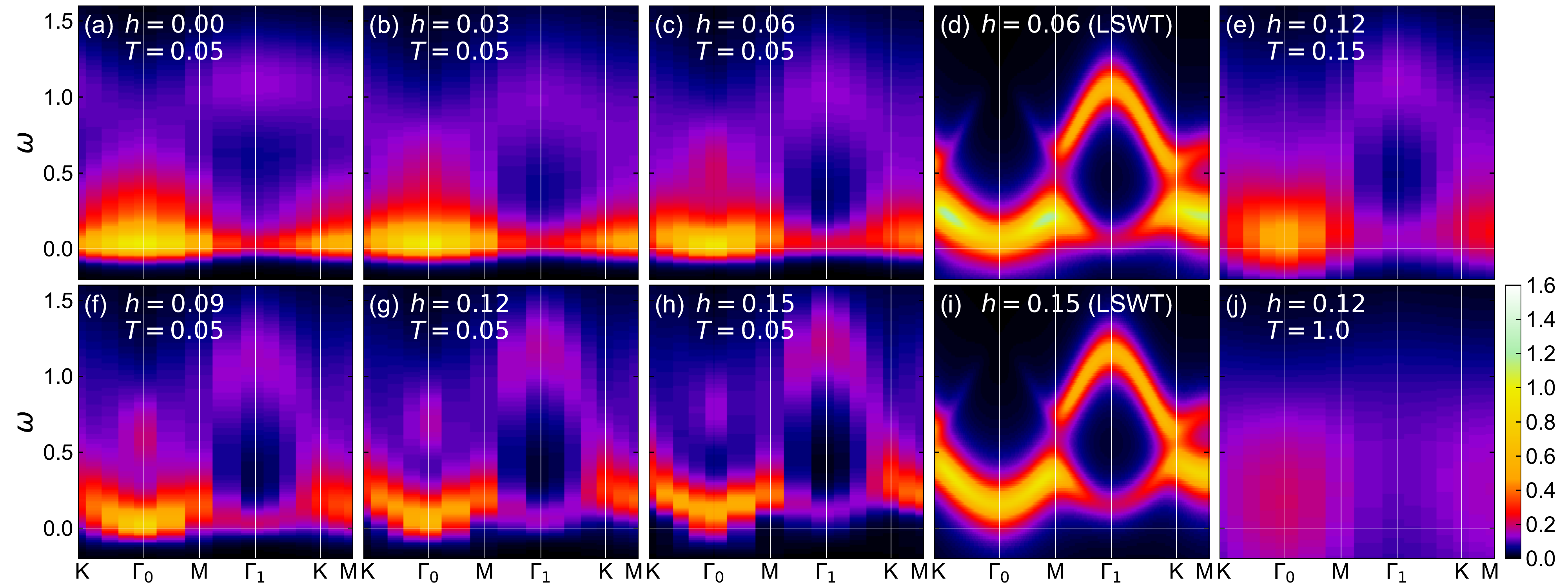}
    \caption{ \label{fig:fig3}
    The dynamical spin structure factor $S(\mathbf{q}, \omega)$ at $T=0.05$ for (a) $h=0.00$, (b) $h=0.03$ (c) $h=0.06$, (f) $h=0.09$, (g) $h=0.12$, (h) $h=0.15$, and at $h=0.12$ for (e) $T=0.15$ and (j) $T=1.0$~\cite{supp}. 
	(b), (c), (e)-(h), and (j) are calculated by the cluster-based CTQMC method, while (a) is obtained by the CDMFT+CTQMC method used in Ref.~\cite{Yoshitake2016,Yoshitake2017a}.
    For comparison, the results obtained by the LSWT are shown for (d) $h=0.06$ and (h) $h=0.15$. 
    The LSWT results are broadened by the smearing factor of $0.1J$. 
    The data are plotted along the symmetric lines shown in Fig.~\ref{fig:fig1}(b). 
    }
\end{figure*}

The results in Fig.~\ref{fig:fig2} suggest that while increasing $h$ in the fractional PM regime, the excitations of the system change from the fractional Majorana fermions and $Z_2$ fluxes to the conventional magnons across the crossover at $T'_L$. 
The magnetic field $h$ hybridizes the two types of fractional excitations [see Eq.~(\ref{eq:H_Maj})], and finally, wipes out the fractionalization by confine them in the forced FM state. 
We investigate such crossover behavior by calculating the dynamical spin structure factor $S(\mathbf{q},\omega)$, which is proportional to the cross section in inelastic neutron scattering experiments. 
The results at $T=0.05$ are shown in Fig.~\ref{fig:fig3}. 
At $h=0$, $S(\mathbf{q},\omega)$ shows incoherent features originating from the fractionalization~\cite{Knolle2014,Yoshitake2016,Yoshitake2017a,Yoshitake2017b}; 
the relatively large intensity at low energy mainly comes from the excitations of the $Z_2$ fluxes, while the broad intensity at high energy are mostly from the itinerant Majorana fermions, as shown in Fig.~\ref{fig:fig3}(a). 
Both features show weak $\mathbf{q}$ dependences, reflecting the extremely short-range spin correlations~\cite{Baskaran2007}. 
When switching on $h$, the fractional spectrum does not show a significant change in the fractional PM region, as shown in Figs.~\ref{fig:fig3}(b) and \ref{fig:fig3}(c). 
With further increasing $h$ across $T'_L$, however, the spectrum is modulated to show dispersive features, as shown in Figs.~\ref{fig:fig3}(f)-\ref{fig:fig3}(h). 
For comparison, we show the magnon dispersions calculated by the linear spin-wave theory (LSWT) for the forced FM state at $h=0.06$ and $h=0.15$ in Figs.~\ref{fig:fig3}(d) and \ref{fig:fig3}(i), respectively~\cite{McClarty2018,Joshi2018,note2}. 
Although the spectra at $h=0.06$ in Figs.~\ref{fig:fig3}(c) and \ref{fig:fig3}(d) look rather different, those at $h=0.15$ in Figs.~\ref{fig:fig3}(h) and \ref{fig:fig3}(i) are much closer to each other. 
We note that while the center of mass of the Majorana band in the weak-field fractional PM state is in the same energy scale for that of the upper magnon band in the high-field forced FM state, the lower magnon band energy increases as $h$ and exceeds the $Z_2$ flux gap for $h \gtrsim h_c$~\cite{supp}. 
The latter behavior is consistent with the extended fractional PM state in Fig.~\ref{fig:fig2}; the lower-energy fractional excitations than that of the lower magnon band contributes to the entropic stabilization of the fractional PM state while raising $T$ from the forced FM state. 

Thus, our results demonstrate how the fractional excitations cross over to the conventional magnon excitations: 
The incoherent features inherent to the itinerant Majorana fermions and the localized $Z_2$ fluxes are reorganized through the hybridization into the dispersive magnon bands by increasing $h$. 
Similar crossover is also driven by varying $T$ in a field. 
This is demonstrated in Fig.~\ref{fig:fig3} for $h=0.12$. 
The magnon-like behavior [Fig.~\ref{fig:fig3}(g)] changes into the incoherent one [Fig.~\ref{fig:fig3}(e)], and finally becomes featureless in the conventional PM [Fig.~\ref{fig:fig3}(j)] while raising $T$~\cite{supp}. 

\begin{figure}[t]
    \includegraphics[width=0.75\columnwidth,clip]{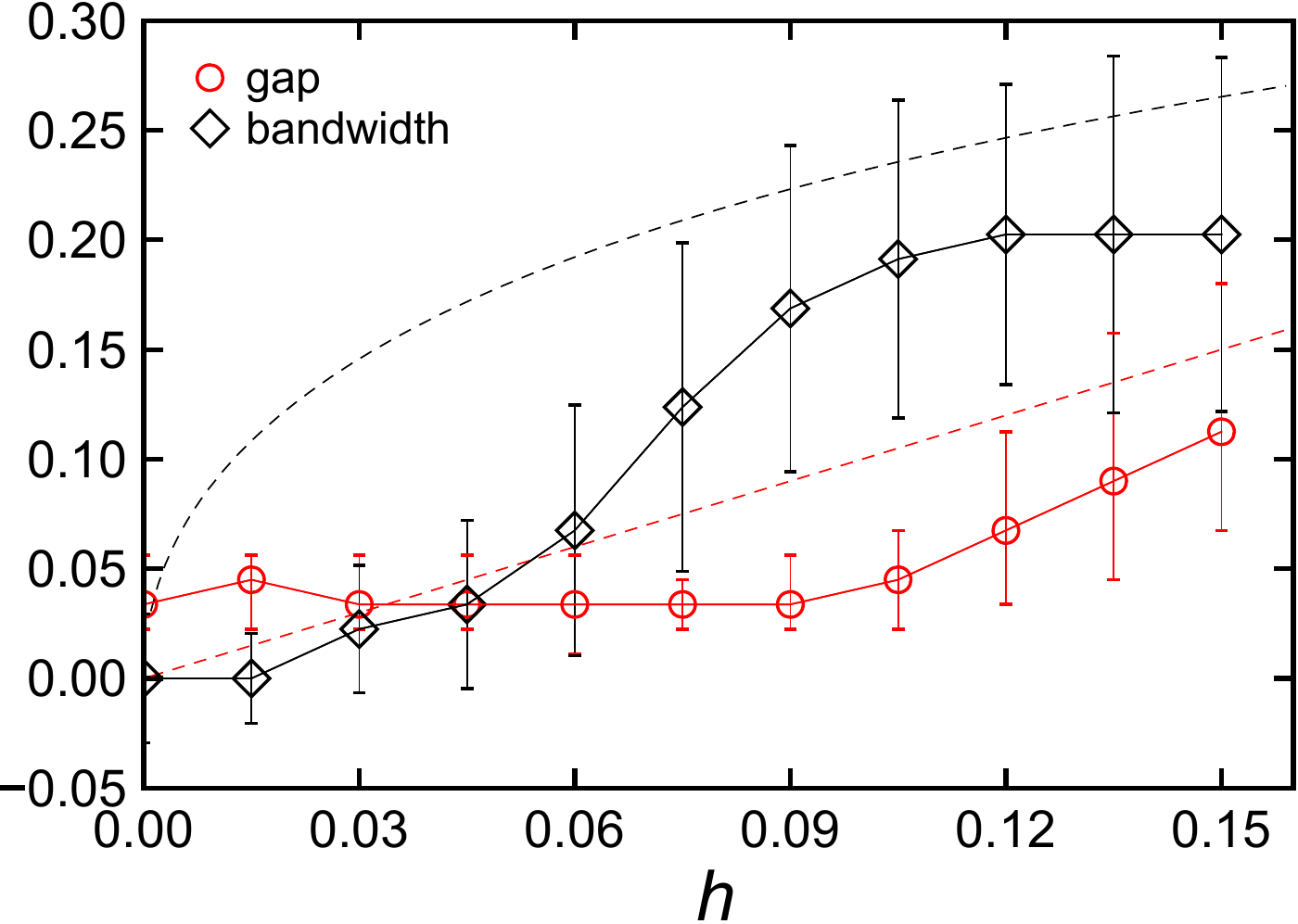}
    \caption{ \label{fig:fig4}
    Magnetic field dependences of the gap and bandwidth, measured by the lowest energy of the peak of $S(\mathbf{q},\omega)$ at $\Gamma_0$ and by the energy difference between the peaks at $\Gamma_0$ and K, respectively~\cite{note3}.
    The dashed lines are the results obtained by the LSWT for the forced FM state. 
    }
\end{figure}

We analyze the confinement-deconfinement behavior more quantitatively, by comparing the gap and bandwidth of the dispersions with the LSWT results. 
The comparison is shown in Fig.~\ref{fig:fig4}. 
We measure the gap by the lowest energy of the peak at $\Gamma_0$, while the bandwidth for the lowest branch along the $\Gamma_0$-K line defined by the energy difference between the peaks at $\Gamma_0$ and K. 
The results show that the gap is almost unchanged and the bandwidth shows a slow change in the fractional PM region, but both rapidly approach the LSWT results through the crossover to the forced FM region; the remaining small deviations might be ascribed to the finite-$T$ effect.

Starting from the fractional quasiparticle picture, the crossover from the fractional PM to the forced FM is driven by quantum many-body effects between the quasiparticles. 
Meanwhile, in the conventional magnon picture, the magnons are multiply generated and interact with each other while approaching the crossover from high $h$. 
Therefore, the excitation spectrum in the crossover region is hard to describe by either picture. 
Recently, several theoretical studies have been done for this challenging issue, but they were thus far limited to an approximate theory~\cite{Banerjee2018} and the exact diagonalization of small clusters typically with 24 spins~\cite{Yadav2016,Winter2017b,Winter2018,Hickey2019}. 
Our results in Figs.~\ref{fig:fig3} and \ref{fig:fig4} provide the unbiased numerical data with much higher resolution in both energy and wave number compared to the previous studies, despite the lack of low-$T$ data due to the negative sign problem. 
A remarkable finding is that the incoherent spectrum well described by the fractional quasiparticles rather than the multiple magnons persists far beyond $h_c$ at finite $T$.


To summarize, we have investigated the crossover between the fractional Majorana quasiparticles and the conventional magnons for the Kitaev model in a magnetic field. 
By developing the cluster-based CTQMC method, we clarified the $h$-$T$ phase diagram and identified the wide fractional PM region distinguished from the asymptotic QSL, forced FM, and conventional PM states. 
We found that the fractional PM region is characterized by the enhancement of the NMR relaxation rate. 
Moreover, we revealed the dynamical spin structure factor retains the unconventional incoherent spectrum in the fractional PM region, while it is modulated to acquire dispersive features through the crossover to the forced FM region. 
The crossover is a confinement of the fractional quasiparticles into magnons, which has been studied, e.g., in quasi-one-dimensional magnets~\cite{Zheludev2002,Lake2009,Coldea2010} and frustrated magnets~\cite{Sachdev1992,Zheng2006,Ito2017}. 
Our unbiased results obtained in two dimensions will be helpful for deeper understanding of the confinement-deconfinement phenomena. 
Furthermore, while our analysis has been limited to the pure Kitaev model, our results offer the firm reference for understanding of interesting behaviors found in the Kitaev candidate materials where the Kitaev interaction is predominant. 
Last but not least, the present study has focused on the FM case, but recently the AFM Kitaev model has attracted much attention from the possibility of an intermediate phase in the magnetic field~\cite{Zhu2018,Gohlke2018,Nasu2018,Ronquillo2019,Hickey2019}. 
The extension of our analysis to the AFM case remains as an interesting future issue. 

\begin{acknowledgments}
J.N. acknowledges the support of Leading Initiative for Excellent Young Researchers in MEXT. 
This work is supported by Grant-in-Aid for Scientific Research under Grant No.~JP16H02206, JP18H04223, JP18K03447, and JP19K03742, and by JST CREST (JP-MJCR18T2). 
Parts of the numerical calculations were performed in the supercomputing systems in ISSP, the University of Tokyo.
\end{acknowledgments}


\begin{thebibliography}{999}
\bibitem{note1} To avoid singular behavior at low magnetic field, we introduce a constant $\alpha$ in the perturbed terms as $-J_\gamma(S_j-\alpha)(S_k-\alpha)$; we take $\alpha=0.1$.
\bibitem{note2} Note that the LSWT results contain only the transverse components to $\mathbf{h}$, while the CTQMC ones include the longitudinal component as well.
\end{thebibliography}
\end{document}